\documentclass{ptapap}
\usepackage{color}

\usepackage{soul}

\usepackage[colorlinks,citecolor=blue,linkcolor=blue,urlcolor=blue]{hyperref}
\usepackage{bm}

\author{Krzysztof Nalewajko}[CAMK]
\affil[CAMK]{Nicolaus Copernicus Astronomical Center, Polish Academy of Sciences\\
Bartycka 18, 00--716 Warszawa, Poland\\
{knalew@camk.edu.pl}}

\title{Relativistic magnetic reconnection in application to gamma-ray astrophysics}
\headtitle{Relativistic magnetic reconnection}

\begin{document}

\maketitle

\begin{abstract}
Cosmic sources of gamma-ray radiation in the GeV range are often characterized by violent variability, in particular this concerns blazars, gamma-ray bursts, and the pulsar wind nebula Crab.  Such gamma-ray emission requires a very efficient particle acceleration mechanism.  If the environment, in which such emission is produced, is relativistically magnetized (i.e., that magnetic energy density dominates even the rest-mass energy density of matter), then the most natural mechanism of energy dissipation and particle acceleration is relativistic magnetic reconnection.  Basic research into this mechanism is performed by means of kinetic numerical simulations of various configurations of collisionless relativistic plasma with the use of the particle-in-cell algorithm.  Such technique allows to investigate the details of particle acceleration mechanism, including radiative energy losses, and to calculate the temporal, spatial, spectral and angular distributions of synchrotron and inverse Compton radiation.  The results of these simulations indicate that the effective variability time scale of the observed radiation can be much shorter than the light-crossing time scale of the simulated domain.
\end{abstract}

\section{Introduction}

The high-energy gamma-ray sky (0.1-10 GeV), as observed by the Fermi Large Area Telescope \citep{2009ApJ...697.1071A}, is dominated by two classes of sources: blazars and pulsars. Gamma-ray pulsars are concentrated along the Galactic plane, while extragalactic blazars are distributed uniformly. Both of these sources are strongly variable, and their gamma-ray emission is non-thermal, indicating efficient mechanisms of particle acceleration.

Gamma-ray variability has been observed in certain blazars and radio galaxies on time scales of several minutes, much shorter than the light crossing time of their supermassive black holes (typically a few hours), e.g.: PKS~2155-304 \citep{2007ApJ...664L..71A}, Mrk~501 \citep{2007ApJ...669..862A}, PKS~1222+216 \citep{2011ApJ...730L...8A}, IC~310 \citep{2014Sci...346.1080A}, 3C~279 \citep{2016ApJ...824L..20A}. It has been argued that such short variability time scales require a highly localized dissipation mechanism, not directly related to variations in the jet structure induced at the central black hole. In addition, such rapid variations observed at very high gamma-ray luminosity impose a potential problem of intrinsic absorption of the gamma-ray radiation in a photon-photon pair creation process. Such absorption can be avoided by postulating very high Doppler or Lorentz factor $\mathcal{D} \sim \Gamma \sim 100$ \citep{2008MNRAS.384L..19B}. In the case of luminous quasars, like PKS~1222+216 and 3C~279, additional absorption of gamma rays can be expected at subparsec scales due to the external radiation field that includes broad emission lines and direct accretion disk radiation.

Relativistic magnetic reconnection has been proposed as a solution to these challenges in the form of the minijets model, in which reconnection produces additional relativistic bulk outflows in the jet co-moving frame, increasing the effective Doppler factor \citep{2009MNRAS.395L..29G}. A semi-analytical model of minijets has been applied directly to the case of PKS~2155-304 \citep{2011MNRAS.413..333N}. However, that model was highly simplified, and over the last several years numerical simulations showed that relativistic magnetic reconnection is a much more complex phenomenon.

Understanding of magnetic reconnection has been developing slowly since the first ideas were formulated in the 1950s in the context of solar physics. Analytical models have difficulty in describing the reconnection process in detail, as it necessarily involves plasma physics beyond the standard MHD or force-free regimes. Kinetic numerical simulations, in particular the particle-in-cell (PIC) algorithm, are our best tools for studying magnetic reconnection in both non-relativistic and relativistic regimes.

Particle acceleration in relativistic reconnection has been studied with PIC simulations since the work of \cite{2001ApJ...562L..63Z}. At that time, it was not even clear whether relativistic reconnection is an efficient dissipation mechanism, as solutions based on smooth Sweet-Parker current layers predicted very low reconnection rates (slow inflow velocities, and hence weak electric fields). It has been known that long current layers are prone to tearing-mode instability, which produces chains of plasmoids, but PIC simulations were necessary to demonstate that formation of plasmoids accelerates the reconnection rate \citep{2004PhPl...11.1151J}.

With increasing computational power, PIC simulations showed that relativistic reconnection is very efficient in accelerating particles, producing power-law particle energy distributions with indices $N(\gamma) \propto \gamma^{-p}$ approaching $p \to 1$ in the limit of relativistic background magnetization $\sigma_0 = B_0^2/(4\pi w_0) \gg 1$, where $w_0$ is relativistic enthalpy including the rest-mass energy \citep{2014ApJ...783L..21S,2014PhRvL.113o5005G,2016ApJ...816L...8W}. Recent simulations of Harris layers with open boundaries showed how reconnection can operate as a steady-state mechanism producing plasmoids of specific size distribution accelerated to relativistic bulk motions \citep{2016MNRAS.462...48S}.
Introduction of synchrotron radiative losses to the PIC algorithm \citep{2013ApJ...770..147C} allowed to study particle acceleration under severe radiative losses, with direct application to the gamma-ray flares from the Crab Nebula \citep{2011Sci...331..736T}, which are interpreted in terms of synchrotron emission exceeding the radiation reaction photon energy limit of $\sim 100\;{\rm MeV}$.
Finally, alternative initial conditions are being explored, i.e., ``ABC fields'' that allow in addition to study the formation and dynamics of current layers \citep{2016ApJ...826..115N,2016ApJ...828...92Y,2017JPlPh..83f6302L}.

One of the most interesting properties of radiation produced by relativistic magnetic reconnection is its rapid variability. This is illustrated with the example of 2-dimensional simulation of a Harris layer described in detail in \cite{2015ApJ...815..101N}.
These results have been obtained by us from numerical simulations performed with the PIC code Zeltron \citep{2013ApJ...770..147C} created by Beno{\^i}t Cerutti\footnote{http://benoit.cerutti.free.fr/Zeltron/}.
The space-time diagrams of synchrotron emissivity and the corresponding lightcurves are presented here for the first time.

\begin{figure}[ht]
\centering
\includegraphics[width=0.7\textwidth]{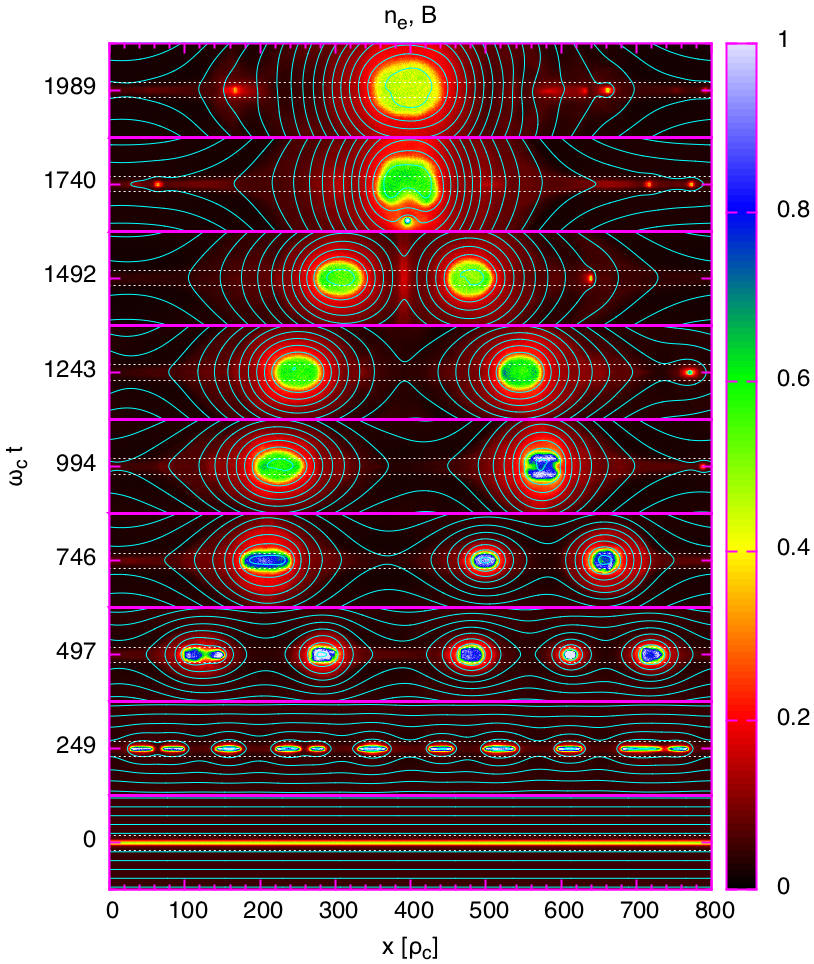}
\caption{Snapshots (x,y-maps) from the evolution of Harris layer undergoing tearing instability and hierarchical mergers of plasmoids. Color scale indicates the particle (electron-positron) density $n_{\rm e}$, and cyan lines indicate the magnetic field lines. Simulation time for each panel is given on the left side. The dashed white lines mark the regions, from which the x-profiles used to build the (x,t) space-time diagrams were extracted.}
\label{fig_xy}
\end{figure}

\section{Rapid variability of emission produced during relativistic reconnection}

A Harris layer is defined as \citep{1962NCim...23..115H,2003ApJ...591..366K}:
\begin{eqnarray}
B_{\rm x}(y) &=& B_0\tanh(y/\delta)\,,
\\
n_{\rm e}(y) &=& n_{e,0}\cosh^{-2}(y/\delta)\,,
\end{eqnarray}
with equilibrium provided by tuning the temperature (pressure) and drift velocity (current density) of the particles in order to match the magnetic field gradient.

Fig.\,\ref{fig_xy} shows the spatial distribution of plasma and magnetic fields along a region centered on initially uniform Harris layer for several simulation times. We can see the formation of multiple plasmods due to the tearing-mode instability and their subsequent evolution. In order to reveal the temporal evolution of plasmoids in greater temporal detail, we extracted 1-dimensional profiles of various plasma parameters that were combined into 2-dimensional space-time diagrams.

\begin{figure}[ht]
\includegraphics[width=0.49\textwidth]{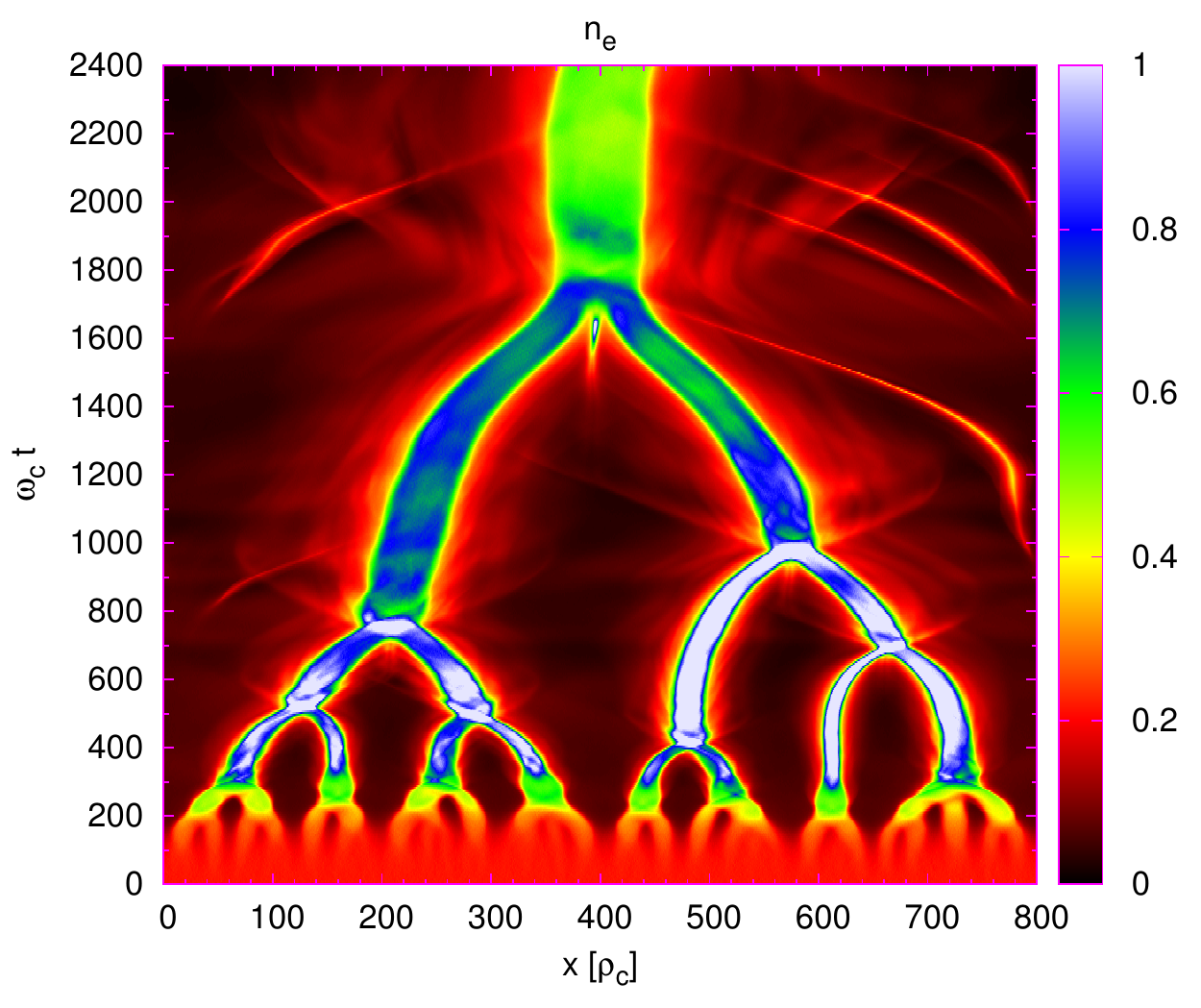}
\includegraphics[width=0.49\textwidth]{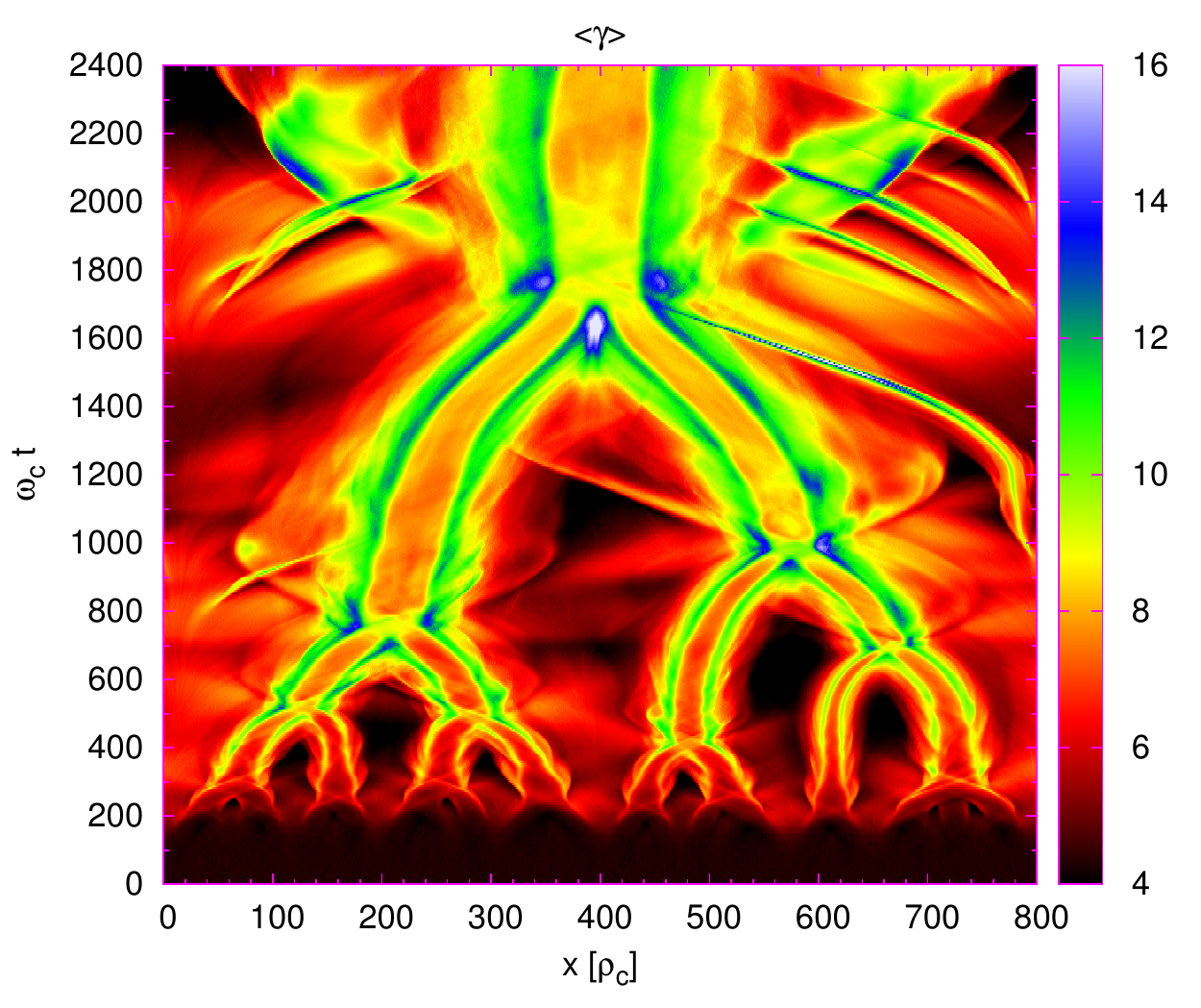}
\caption{Space-time diagrams of particle density $n_e$ (left) and particle mean energy $\left<\gamma\right>$ (right) for the same simulation as shown in Fig.\,\ref{fig_xy}.}
\label{fig_xt}
\end{figure}

Fig.\,\ref{fig_xt} shows two examples of such space-time diagrams: for particle density $n_{\rm e}$ and for mean particle energy $\left<\gamma\right>$ (see \citealt{2015ApJ...815..101N} for more examples). Plasmoids can be seen to consists of two main parts: dense cool cores, and hot dillute layers.
The synchrotron radiation power scales with both particle density, particle energy, and magnetic field strength. It turns out that the total synchrotron radiation power is concentrated along the hot layers, rather than the dense cores.

\begin{figure}[ht]
\includegraphics[width=0.49\textwidth]{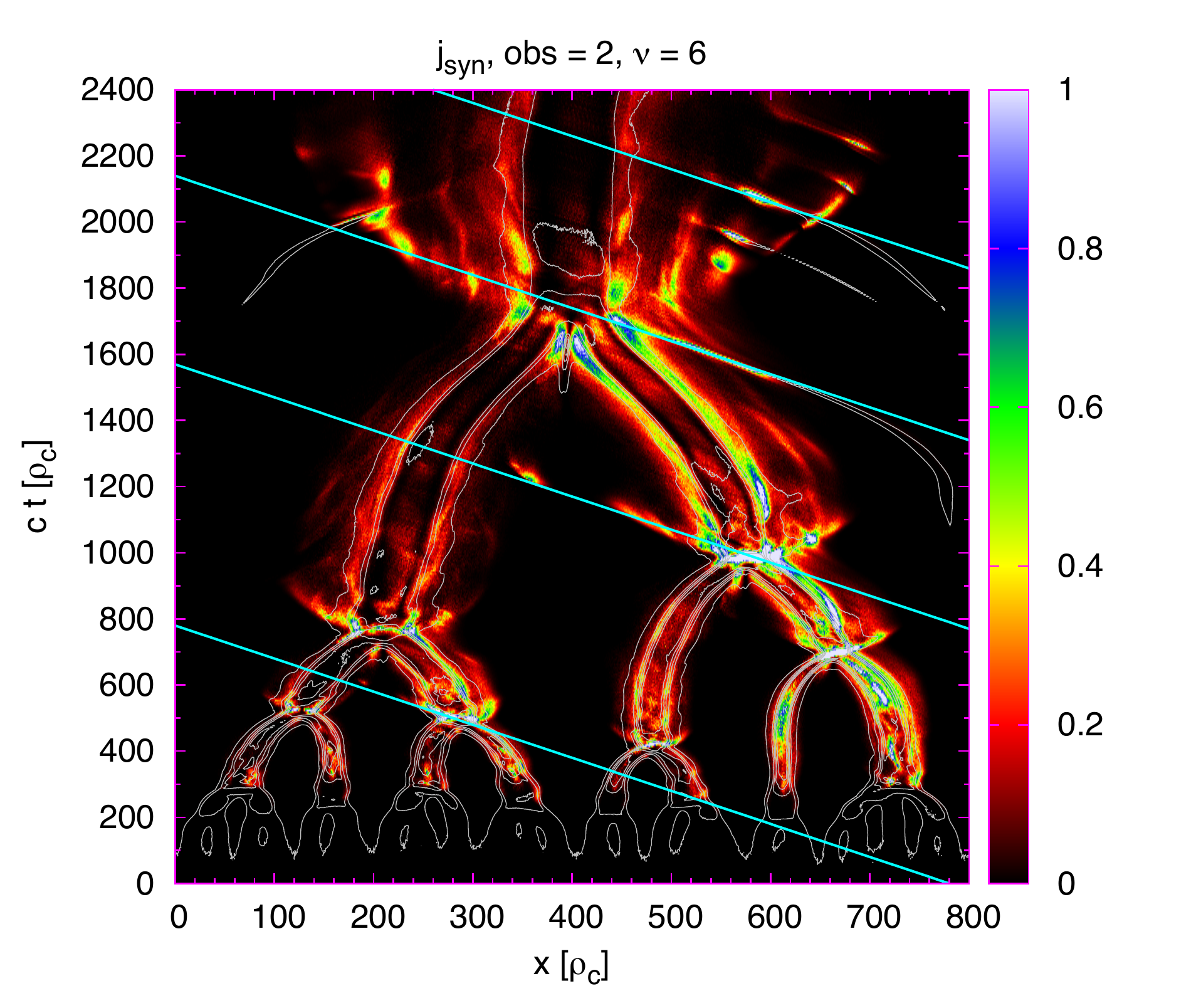}
\includegraphics[width=0.49\textwidth]{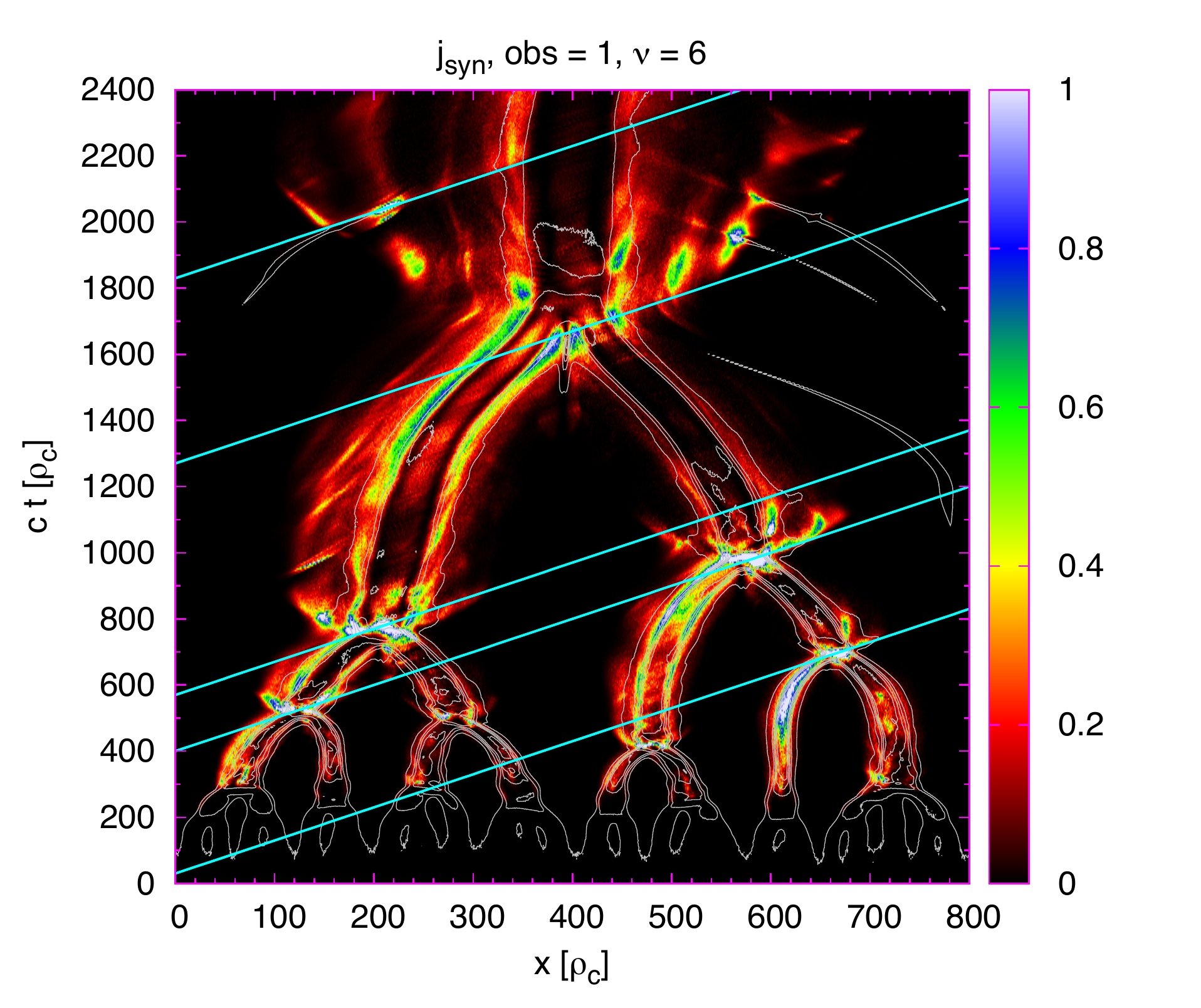}
\includegraphics[width=0.49\textwidth]{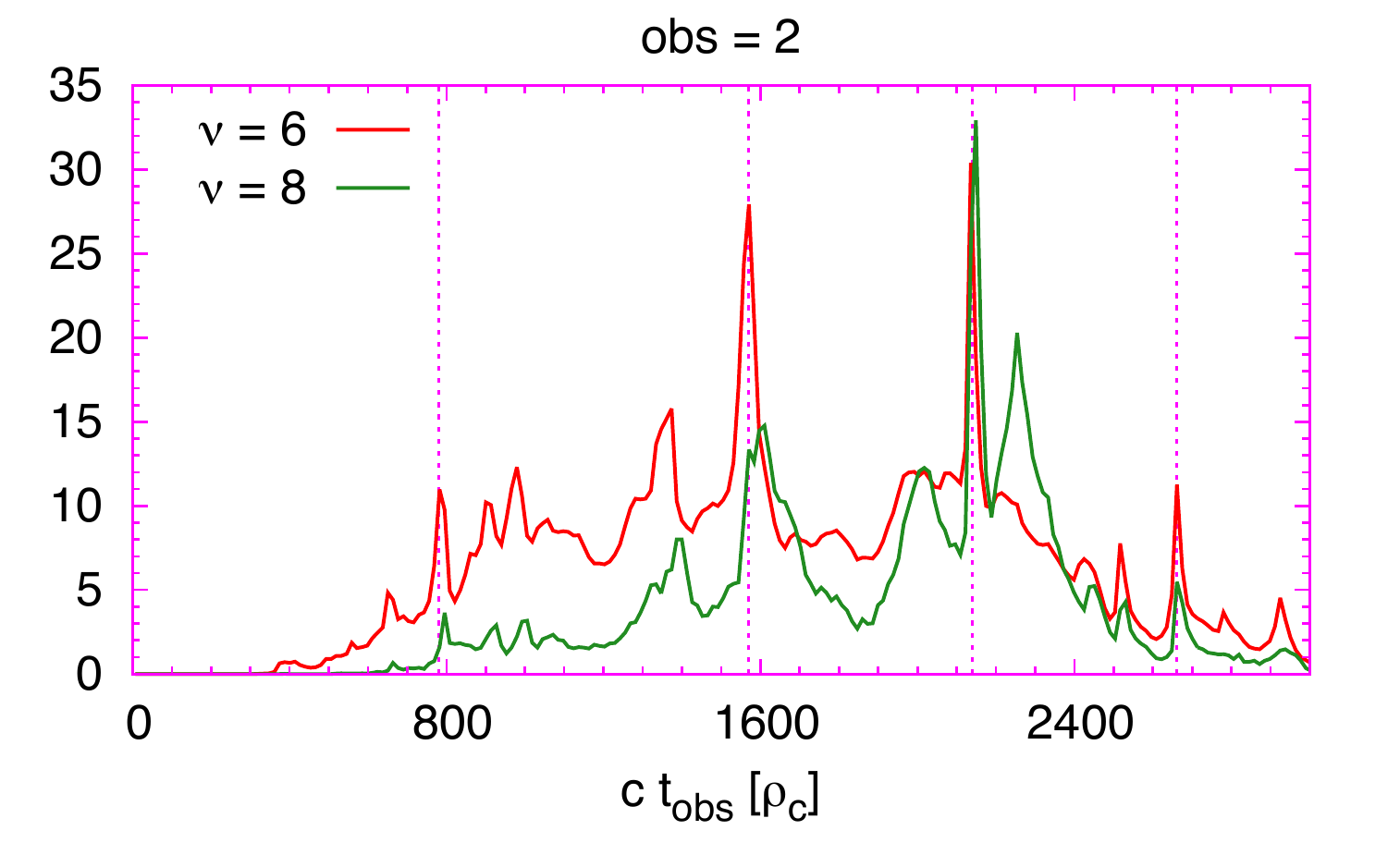}
\includegraphics[width=0.49\textwidth]{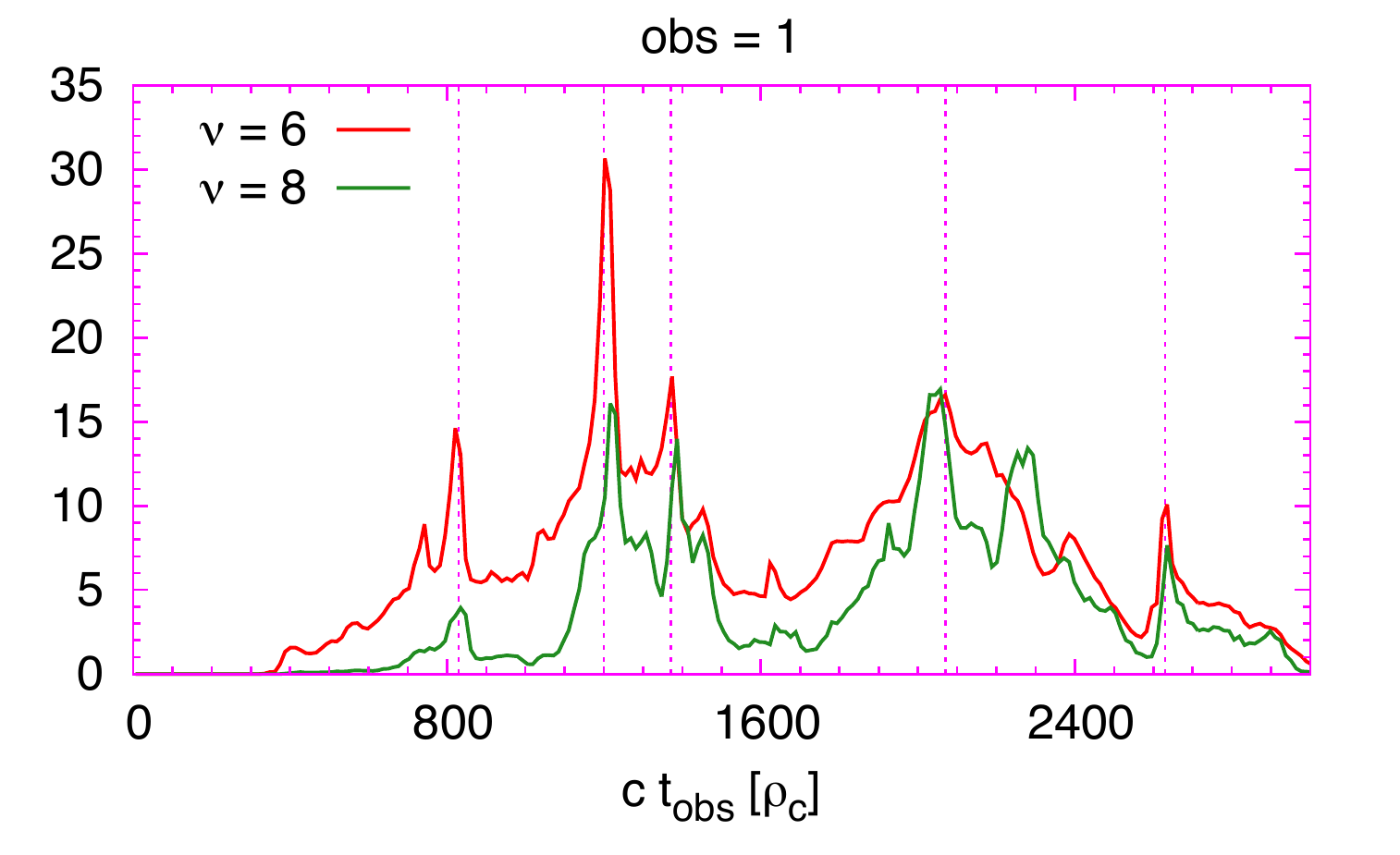}
\caption{Top row: space-time diagrams of synchrotron emissivity directed towards two observers: Observer 1 (right panels) is located at $+x$, Observer 2 (left panels) is located at $-x$. Bottom row: lightcurves expected for Observers 1 and 2 for two frequency bands. Vertical dashed lines marking several flares seen by each observer correspond to the light cones (cyan lines) on the space-time diagrams above.}
\label{fig_lc}
\end{figure}

Fig.\,\ref{fig_lc} shows synchrotron emissivity calculated for two opposite observers: Observer 1 at $+x$ and Observer 2 at $-x$. It can be seen that emissivity, like the total power, is concentrated along the hot plasmoid layers. However, the difference between emissivity distributions directed towards opposite observers demonstrates significant anisotropy of the radiation. Observer 1 detects more radiation from the plasmoids propagating to the right, while the plasmoids propagating to the left contribute very little to the observed emission.

Each space-time diagram of emissivity can be converted into the observed
lightcurve by collecting radiation along the light cones indicating fixed observation times $t_{\rm obs}$.
In the bottom row of Fig.\,\ref{fig_lc}, we show lightcurves that correspond exactly to the space-time diagrams of synchrotron emissivity presented in the upper row.
Several characteristic observation times are indicated, i.e., moments of major observed flares/spikes.
Each dashed vertical line in the lightcurves has a corresponding light cone in the space-time diagram. These light cones are essential in order to locate the emission event responsible for each observed flare. We can see that most of such light cones pass through a plasmoid merger event. We can thus associate sharp radiation flares with plasmoid mergers.

The lightcurves shown in Fig.\,\ref{fig_lc} indicate that synchrotron radiation produced during relativistic magnetic reconnection can be variable on time scales shorter by order of magnitude from the global light-crossing time scale. Indeed, while the light-crossing time of the simulation domain is $c\, \Delta t_{\rm obs} = 800\, \rho_{\rm c}$, the observed flares are clearly shorter than $100\,\rho_{\rm c}$. Two reasons for such rapid variability have been suggested \citep{2012ApJ...754L..33C}: spatial bunching vs. sweeping beams. Spatial bunching corresponds to highly localized emitting regions, our space-time diagrams of synchrotron emissivity suggest emitting events localized both in space and in time. Sweeping beams result from a highly anisotropic distribution of energetic particles that are forced to gyrate collectively in magnetic field. Relativistic reconnection results in particle anisotropy that is strongly energy-dependent, this effect is known as kinetic beaming \citep{2012ApJ...754L..33C}, and is distinct from the bulk Doppler beaming (relativistic jet). Analysis of the time evolution of angular distribution of radiation revealed that these kinetic beams are sweeping across certain observers. It has been difficult to resolve this dilemma, it seems that both spatial bunching and sweeping beams are important in modulating the observed radiation signals.

Qualitatively similar results were obtained from analysis of radiation signatures of 2-dimensional magnetostatic structures called ``ABC fields'' \citep{2016ApJ...826..115N,2016ApJ...828...92Y}. These structures are defined as:
\begin{eqnarray}
B_x(x,y,z) &=& B_0\left[\sin(\alpha z)+\cos(\alpha y)\right]\,,
\\
B_y(x,y,z) &=& B_0\left[\sin(\alpha x)+\cos(\alpha z)\right]\,,
\\
B_z(x,y,z) &=& B_0\left[\sin(\alpha y)+\cos(\alpha x)\right]\,,
\end{eqnarray}
where $\alpha = 2\pi k/L$. Initial equilibrium is provided by smoothly distributed current density $\bm{j} = (kc/L)\bm{B}$ obtained by shaping the local momentum distributions of uniformly spaced particles. For $k > 1$, this configuration is unstable to coalescence modes that result in formation of dynamical current layers where magnetic reconnection and particle acceleration take place. Most of the high-frequency synchrotron radiation is produced when energetic particles leave the current layers and begin to interact with strong perpendicular magnetic fields. Once again, we find evidence for both spatial bunching and beam sweeping taking place simultaneously, the observed lightcurves show spikes on time scales order-of-magnitude shorter than the light-crossing time of the simulation domain.

The effective reduction factor for variability time scale of radiation produced during relativistic magnetic reconnection remains unknown. It can be defined formally as $f_\gamma = R_{\rm diss}/(ct_{\rm obs})$,
where $R_{\rm diss}$ is the characteristic radius of the dissipation region.
In \cite{2016ApJ...824L..20A}, we suggested that it can be of the order of $f_\gamma \sim 10-100$, however, further investigation is necessary.

\acknowledgements{The author is grateful to Beno{\^i}t Cerutti, Dmitri Uzdensky, Gregory Werner, Mitchell Begelman, Jonathan Zrake, Yajie Yuan, Roger Blandford, and Martyna Chru{\'s}li{\'n}ska for collaboration on PIC simulations of relativistic reconnection. This work was supported by the Polish National Science Centre grant No. 2015/18/E/ST9/00580.}

\bibliographystyle{ptapap}
\bibliography{knalew}

\end{document}